\documentclass{vgtc}                          

\graphicspath{{figures/}{pictures/}{images/}{./}} 

\usepackage{times}                     

\usepackage{tabu}                      
\usepackage{booktabs}                  
\usepackage{lipsum}                    
\usepackage{mwe}                       

\usepackage{mathptmx}                  

\onlineid{0}

\vgtccategory{Research}

\vgtcinsertpkg

\title{Immersive Focus+Context Techniques to Assist in Interpretation of High Density Forest Point Clouds}

\author{
    \begin{tabular}{ccc}
        \begin{tabular}[t]{c} Spencer O'Keeffe\thanks{e-mail: spencer.okeeffe@mymail.unisa.edu.au}\\
        \scriptsize University of South Australia \end{tabular} &
        \begin{tabular}[t]{c} Bruce H Thomas\thanks{e-mail: bruce.thomas@unisa.edu.au}\\
        \scriptsize University of South Australia \end{tabular} &
        \begin{tabular}[t]{c} Jim O'Hehir\thanks{e-mail: jim.ohehir@unisa.edu.au}\\
        \scriptsize University of South Australia \end{tabular} \\
        \begin{tabular}[t]{c} Jan Rombouts\thanks{e-mail: jrombouts@onefortyone.com}\\
        \scriptsize OneFortyOne Plantations Pty Ltd \end{tabular} &
        \begin{tabular}[t]{c} Michelle Balasso\thanks{e-mail: michelle.balasso@onefortyone.com}\\
        \scriptsize OneFortyOne Plantations Pty Ltd \end{tabular} &
        \begin{tabular}[t]{c} Andrew Cunningham\thanks{e-mail: andrew.cunningham@unisa.edu.au}\\
        \scriptsize University of South Australia \end{tabular}
    \end{tabular}
}

\abstract{
Virtual Reality (VR) is changing how we interact with complex spatial datasets.
VR has the potential within forestry to help analyse remotely sensed spatial data for inventory management. However, challenges of scale and legibility of dense forest point clouds must be addressed. To do so, we present three VR Focus+Context (F+C) techniques: \userfocustag, \remotefocustag, and \rayfocustag. 
These techniques present point cloud data using a constrained high-detail focus within a broader abstracted context representation, manipulated using egocentric and exocentric interactions.
We conducted two user studies.
The first informed the design of the F+C techniques. 
The results from the first study indicate that VR significantly outperforms desktop displays in basic analysis of forestry point clouds, with improvements to both time and accuracy. 
The second study evaluated the performance of the resulting F+C techniques in a task analogous to real-world forestry operations.
Our F+C techniques significantly outperformed current approaches, confirming their viability for facilitating detailed and effective forest assessments.
These findings indicate that VR has significant advantages for spatial data analysis tasks in forestry, making a strong argument for its adoption in future analytical frameworks. 
The F+C visualisation techniques we introduced show considerable promise for analysis of high density point clouds, opening new avenues for development in immersive analytics.
  Supplemental materials (code, scans, results) are available at \url{https://osf.io/hqc9u}.
}


\teaser{
  \centering
  \includegraphics[width=\linewidth, alt={A view of a city with buildings peeking out of the clouds.}]{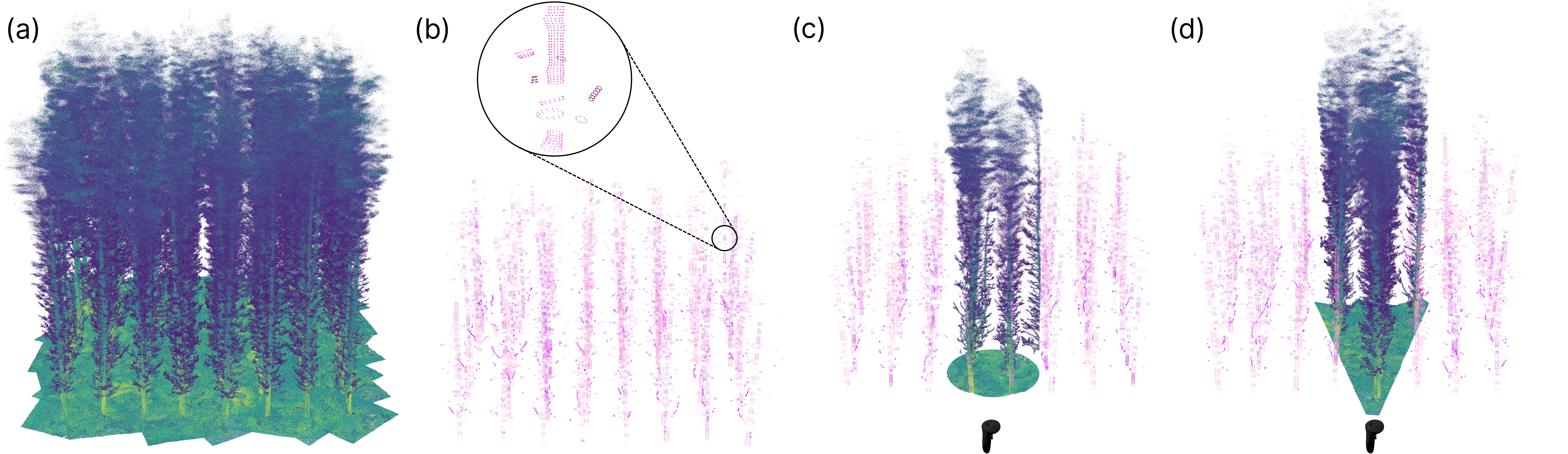}
  \caption{%
  	Focus+Context techniques for dense forest point cloud data: \hl{\textbf{(a)} An example of a dense forest scan. \textbf{(b)} The AI derived abstract tree model, with a callout depicting a single tree. Perspective views for the \textbf{(c)} lantern and \textbf{(d)} wedgelight techniques.}%
  }
  \label{fig:hero}
}

\graphicspath{{figs/}{figures/}{pictures/}{images/}{./}} 

\usepackage{tabu}                      
\usepackage{booktabs}                  
\usepackage{lipsum}                    
\usepackage{mwe}                       

\usepackage{mathptmx}                  

\usepackage{soul}
\usepackage{color}
\usepackage{multirow}

\usepackage{siunitx}
\usepackage{caption}

\UseRawInputEncoding

\usepackage{ifthen}
\newcommand{\contextdatasetdesc}[1]{%
    \ifthenelse{\equal{#1}{upper}}%
    {Synthetic}%
    {synthetic}%
}
\newcommand{\userfocustag}{Wedgelight}
\newcommand{\remotefocustag}{Remote Lantern}
\newcommand{\rayfocustag}{Ray Lantern}
\newcommand{\controltag}{Control}

\newcommand{\sigleveltwo}{**\(p < 0.01\).}
\newcommand{\siglevelthree}{***\(p < 0.001\).}
\newcommand{\appendixtag}{Appendix 1}

\renewcommand\hl[1]{#1}

\begin{document}


\firstsection{Introduction}

\maketitle

In this paper, we present a set of Virtual Reality (VR) visualisation techniques designed to improve the analysis of dense forest point clouds.
Our techniques are based on a Focus+Context (F+C) approach~\cite{kosara_interaction_2003}, whereby a user-controlled \textit{focus} is displayed in detail, while the data \textit{context} is displayed at some higher level of abstraction.
Traditional approaches to silviculture are data-informed but imprecise: 
tree selection, the process by which specific trees are marked to be thinned from the stand (a contiguous community of trees) to maximise the stability and growth of the remaining trees, relies on snap judgements by operators in the field.
Similarly, stand assessment, the process of taking a measurement of the forest, traditionally involves measuring a small subset of the trees and extrapolating.
Both processes would benefit from higher precision.

Forest practitioners are increasingly making use of remote sensing data for forest management purposes. 
High density LiDAR or photogrammetry based point clouds are captured to generate high fidelity 3D depictions of the forest being managed~\cite{akay_using_2009}. The objective is to deploy these datasets towards optimisation of forest planning and silviculture.
Increasingly, forest managers are exploring whether the digital representations of forests can be used to reduce the need for field inspections, to transfer traditionally field based tasks to the office, providing a conduit to more precise and automated operations. 
For this to be possible forest practitioners need to be able to interact with the digital forest in an effective and ergonomically sound manner, including with the raw data that provide the basis for the construct. 

VR presents an opportunity to improve visualisation and interaction with dense point clouds.
VR displays are inherently immersive~\cite{slater_framework_1997}, and have been shown to enable a higher level of spatial recall and understanding of point clouds than 2D representations~\cite{reichherzer_bringing_2021}, suggesting that VR may be ideal for advanced 3D forestry analytics.
However, the rendering requirements of VR bring strict performance limitations, and the methodology for comfortable and efficient interaction with dense point clouds is still unclear, so designing suitable implementation methods is essential.
To this end, we seek to answer the following: \textit{How do we leverage VR to support dense point cloud tasks such as is needed in forestry management?}

We developed three VR F+C techniques for understanding forest point cloud data: \userfocustag, \remotefocustag\ and \rayfocustag.
The design of these techniques is informed by results of a formative study and feedback from domain experts. Each technique, named following an illumination metaphor, is a combination of an egocentric or exocentric control scheme, and a focus shape.
This set of techniques allow users to assess point clouds through direct manipulation of a high-resolution focus, within the context of a simplified structural visualisation. 
Our specific implementation uses AI derived tree features as a \textit{context} dataset, and the \textit{focus} dataset is a high density point cloud produced by LiDAR scans.
%

In this paper, we present the design of our techniques, beginning with the results of a formative study to examine how VR visualisation can support understanding of forestry data compared to the current standard of desktop displays. 
This initial investigation informed the design of our F+C techniques.
We describe the design of our techniques and the motivating requirements.
We conducted a subsequent user study to evaluate the practical utility of our F+C techniques in an industry context, assessing their performance in a task analogous to a tree thinning operation.
Assessment of user performance on a series of success parameters, movement statistics, and participant feedback were utilised to validate the techniques.
While our assessment focuses on forestry, these visualisation techniques have broader applications in other domains dealing with dense point cloud data.

The key contributions of this paper are the empirical results of a study showing the advantages of VR over Desktop for a set of forestry point cloud analysis tasks;
the design and implementation of a set of novel F+C visualisation techniques for analysing dense forest point clouds in VR; 
shared scan data and study scripts;
and results of a study assessing the performance of the visualisation techniques for the analogue of a real world forestry operation.

\section{Background}

\subsection{Immersive Analytics}

Immersive Analytics \cite{chandler_immersive_2015} is a research area that explores how human computer interaction and immersive display technologies can facilitate complex data analysis.
\hl{Part of the field explores how 3D visualisations can improve assessment of abstract and 3D data }\cite{zhao_metacast_2024}.
\hl{Such visualisations enhance depth perception and presence for users, improving comprehension and analysis} \cite{kraus_immersive_2022, prouzeau_scaptics_2019}.

Immersive analytics tools have been explored for a variety of industrial applications.
Tasks like analysis of multidimensional volumetric datasets for materials analysis \cite{gall_imndt_2021} see advantages in structure and pattern identification, with observations noting that 3D visualisations are highly suited for assessing curved structures, which are otherwise difficult to track and analyse using a 2D cross section. 
\hl{Immersive analytics tools have provided an `accessible and fully immersive comprehensive picture' of physicochemical parameters in biological datasets} \cite{maes_minomics_2018}.
Forestry is not absent in this regard, with explorations in measurement and annotation of field scans in VR \cite{interpine_nz_operational_2020}.

\subsection{VR in a Forestry Data Context}

There have been efforts to explore immersive analytics tools like VR within forestry operations, such as with the aforementioned field scan annotation.
VR displays are fundamentally well suited to analysis of complex spatial forestry data:
3D displays can alleviate the difficulty of analysing complex 3D datasets~{\cite{ware_visualizing_2008}},
and the 3D environment can facilitate rapid identification and selection of features in such data~{\cite{franzluebbers_virtual_2022}}.
Analytical task performance has been shown to vary with level of immersion; increased immersion offers substantial benefits for 3D data and cluster detection~{\cite{kraus_impact_2020}}, which has applications in forestry analytics.

However, dense forest point cloud datasets present  challenges; the volume and area of scans can be on the scale of hectares, producing vast datasets.
The position of trees is often scattered, and neighbouring branches commonly overlap, creating occlusions and identification problems.
Accurately assessing trees requires looking into branch and stem structure, which can be hidden from the outside.
There are also challenges with utilising VR in forestry to its full potential.
In one recent case, Widjojo et al.~\cite{widjojo_using_2021} explored using immersive analytics for forest assessment, looking specifically at inventory assessments of radiata pine data.
Following a user study assessing the tools, they noted that the density of points within a tree, and the segmentation techniques used on the point cloud both impacted assessment performance considerably.
A lack of colour information on points also made tasks more difficult.
They noted that ``achieving a higher density of point clouds in certain areas in a VR-point cloud...is likely to be sufficient to improve task performance in the VR-point cloud''.
A systematic literature review by Perez-Huet \cite{perez-huet_global_2020} further highlights some of these barriers.
The review notes that transitioning visualisations between scales of forested areas is a challenge, as is designing displays with Level of Detail (LOD) to maintain performance, while still emphasising that VR models can be advantageous in creating ``flexible, interactive, and controlled ways to view [forestry data]''.
VR has also seen use as an outreach and engagement tool in the forestry sphere \cite{borda_virtual_2022,lewis_computer-based_2005}, as well as application in training and planning operations \cite{huang_walking_2021,lapointe_using_2000,moreno_interactive_2014}.



Performance limitations make rendering high resolution point clouds in VR a challenge, independent of analytical context.
They can be circumvented by using clouds that do not exceed the upper performance limit for the display device:
for example, Stets et al.~{\cite{stets_visualization_2017}} were able to label the data from a Microsoft Kinect with no need for pre-processing.
For point clouds that go beyond this performance threshold, pre-processing or display techniques are required to display the data comfortably.
Ramirez et al.~{\cite{ramirez_shooting_2020}} use a voxelization algorithm to process their data before displaying it, reducing display complexity while still providing structural scene context.
However, this approach is not suited to tree data for all cases, as without data of sufficiently high resolution the voxelization process is likely to blur the distinction between branch and leaf, stem and crown.
Another approach that sees use in research projects~\cite{ramirez_shooting_2020, franzluebbers_virtual_2022, schutz_real-time_2019} is LOD systems: dividing the scan data into subsets and \hl{rendering each at} a lower resolution based on its proximity to the focal point.
This improves performance, while still facilitating assessment of the whole dataset.
A similar approach is manual selection of area of interest, such as using a cross section to isolate a subset of a point cloud for closer inspection~{\cite{drissi_immersive_2021}}.
Improvements can be gained by small optimisations like frustum culling (unloading the points outside of the users view) which sees use in point cloud interaction tools \cite{garrido_point_2021}.
Alternatively, one can simply compress the dataset by reducing its resolution altogether.
This can be achieved through de-noising and refinement processes that leverage techniques like octree encoding and surface approximation \cite{graziosi_overview_2020,huang_generic_2008}.
Point cloud editors like CloudCompare \cite{girardeau-montaut_cloudcompare_2016} offer tools that can procedurally remove points from a cloud,\hl{ trading quality for size and performance,} but this comes at the risk of sacrificing details if processed over-zealously.
Addressing performance limitations is necessary to optimise forest point cloud analysis in VR.

\subsection{Focus+Context}

A technique that may work to remediate some of these issues is the F+C methodology described by Kosara et al.~\cite{kosara_interaction_2003} succinctly as ``a way to zoom in on parts of a large dataset without losing track of where one is within the whole data''. 
Early work by Furnas~\cite{furnas1981fisheye} conceptualised F+C as a focal point, a degree of interest function and the context, \hl{which has since been generalized }\cite{hauser_generalizing_2006} \hl{and seen a variety of approaches} \cite{cockburn_review_2008}.
A simplification of this F+C method is the \textit{Filtering} approach, whereby instead of distorting or separating the point of interest, additional information is provided only in areas of interest in the dataset.
A conceptual example exists in the form of 3D Magic Lenses \cite{10.1145/237091.237098} as early as 1996, however this render is dependent on the users perspective.
Kalkofen et al. \cite{kalkofen_interactive_2007} developed an immersive F+C visualisation frame work for an Augmented Reality (AR) context.
Their work demonstrated how the ability to focus on important information while still perceiving contextual information in the periphery can reduce clutter while improving spatial comprehension.
As they put it, F+C techniques ``are highly successful in scientific visualisation'', 
suggesting they may be a suitable solution to the challenges of utilising VR in a forestry data context.

















    
    
    
    
    





\section{Study 1: Point Cloud Comprehension on Desktop Versus VR}
\label{study1}

To inform the design of our forestry visualisation techniques, we conducted a formative controlled user study to understand the effects of different display modalities on comprehension of forestry point cloud data.
To this end, we compared VR to traditional desktop displays.
To focus on the effects of stereopsis and structure from parallax independent of display resolution, we controlled for both the display resolution and system input (as described in Section \ref{s1-apparatus}).
The study design was approved by the University Human Research Ethics Committee.


\subsubsection{Task}
To ensure silvicultural validity, the tasks were designed to be analogues to real-world forestry tasks.
Participants were required to answer a set of comprehension questions about forestry point cloud scans, repeated for each display condition.
To maximise the relevance of study results to industry, questions were based on domain expert feedback and real world point cloud scans were used.
(Subsets of the data described in {\ref{data-collection-section}} were used for each question task.)
We utilised 6 types of comprehension question, each with 4 unique variations, totalling 24 tasks:

\textbf{Counting:} the number of trees in a scan comprising 20--30.

\textbf{Branching:} find the tree with the most branches from 3 trees.

\textbf{Tallest:} Identify the tallest tree from 10--25 trees.

\textbf{Cluster:} Identify the tallest group of 3 trees from 3 groups.

\textbf{Split:} Identify the tree with a split stem from 20--25 trees.

\textbf{Mirror:} Identify the two identical trees from 6 trees.


Participants were prompted with the task objective (\textit{ie. Identify the tallest tree}), then allowed unlimited time to inspect the associated point cloud, and finally prompted to make a selection on an answer screen. 
This answer screen allowed users to select an integer from a list, representing a tree count or specific tree ID, depending on the task.
Participants could use a controller to rotate, scale, and translate the dataset during the task to change their perspective.


\subsubsection{Independent Variables}

Independent variables were counterbalanced to avoid effects of fatigue or learning. 
The two independent variables were as follows:

\textbf{Display Condition:} Desktop or Virtual Reality.

\textbf{Question Type:} The different question type (Counting, Branching, Tallest, Tallest Cluster, Split, Mirror).





\subsubsection{Dependent Variables}

The seven study dependent variables were as follows:

\textbf{Correctness:} Binary correct or incorrect indicating if the participant correctly answered the question.

\textbf{Tree Counting Accuracy:} Measured for the Counting task, the distance to the correct answer, calculated as the \hl{relative} difference between the users answer and the true number of trees in the task.

\textbf{Cloud Observation Time:} The time (in seconds) the participant took to assess the point cloud for each task. 

\textbf{Frame-of-Reference Transformation:} Cumulative FoR metrics (translation, rotation, scale, and resets) recorded every 0.1s.

\textbf{Mental Effort:} Response on a nine point Likert scale \cite{likert_technique_1932}.

\textbf{System Usability:} Participants SUS \cite{brooke_sus_1995} questionnaire ratings.

\textbf{Qualitative Feedback:} Preferred Condition for Viewing; Perceived Accuracy, Enjoyment, and Overall, and any additional commentary about the techniques and the study experience.


\subsubsection{Hypothesis}

The hypotheses for the study are as follows:

\textbf{H1} \textit{VR} display will improve user performance over the \textit{Desktop}.

\textbf{H2} Users will prefer the \textit{VR} display over the \textit{Desktop} display.

\textbf{H3} The impact of the \textit{VR} display will vary across task types. 

\subsubsection{Apparatus}
\label{s1-apparatus}
Both study conditions were developed and run through the Unity editor running version 2021.3.16f1 through use of a Dell Precision 7760 Laptop (Win 10, 32 GB RAM, 11th Gen Intel Core i7-11850H, RTX 3080).
The VR condition used a Meta Quest Pro tethered to the laptop via Quest Link.
The Desktop condition was run by extending the laptop display to a \SI{51}{\cm} wide (23 inches diagonal) monitor positioned \SI{70}{\cm} in front of the participant. The display of the monitor was down-sampled to 900 $\times$ 600 pixels to match the effective resolution of the Meta Quest Pro for the equivalent field of view of the monitor (a display width of \SI{51}{\cm} placed \SI{70}{\cm} from the user's eyes has an equivalent FOV of 40$^{\circ}$; 
given the Meta Quest Pro \textit{pixels per degree} of 22PPD, 40 $\times$ 22 $\approx$ 900). 
A standard Xbox 360 Gamepad was used for interacting with the displayed datasets, and for the answer entry system. 
Participants remained seated for the desktop condition. In the VR condition they were invited to either sit or stand, as long as they were consistent.

\subsubsection{Procedure}
After a preliminary introduction and consent form, the study was explained to participants.
Task training was conducted on the desktop first, to allow participants to familiarise themselves with the controls, then repeated in VR.
There was no time limit any task, but users were asked to be as fast and as accurate as possible.
For both conditions, participants completed every task, then a post condition survey.
They concluded with a post study questionnaire.
This resulted in 6 (question types) $\times$ 4 (variations) $\times$ 2 (conditions).

\subsubsection{Participants}
\hl{We recruited 24 participants from university staff and students, (16 male, 8 female) aged 19--37 ($M=28$, $SD=4.6$).}
Each participant was required to have normal or corrected to normal vision.
Eighteen participants were noted as having a background in computer science. The remainder were split across the fields of Architecture, Design, Business, Accounting, Psychology and Environmental Science.
\hl{Participants received a \$20AUD ($\approx$\$13USD) gift card.}


\subsection{Study 1 Results}
\label{s1-results}
We used linear mixed models to explore the influence of most factors on participant performance, as facilitated by R and the package lme4 \cite{bates_fitting_2015}.
The models included fixed effects for display condition and question type, applied to measures of time, correctness, and distance to answer.
Participant variability was accounted for as a random effect.
Statistical significance was assessed using Type II Wald Chi-square tests, with a predetermined alpha level of 0.05.
To analyse participant feedback metrics, the Wilcoxon rank sum test was employed. 
Analysis of participant system preference was conducted using Chi-square goodness-of-fit tests.
Lastly, participant feedback and verbal commentary were recorded and analysed post-hoc to identify recurring themes and capture feedback.
Where question type was found to be a sole significant factor during analysis it was disregarded, as the varied requirements of each question type resulted in vastly different outcomes. We found no interaction effect between question type and any other factors, suggesting the impact of the VR display did not vary across task types.


\subsubsection{Correctness and Tree Counting Accuracy}

We found a significant effect of display condition on \textbf{Correctness}, as evidenced Table~\ref{tab:study1_performance}. Contrast between Desktop ($M = 0.64$, $SD = 0.48$) and VR ($M = 0.70$, $SD = 0.46$) indicated that Desktop was significantly less correct than VR ($t(1139) = -2.588$, \textit{estimate} $= -0.066$, $p = $\textbf{.010}), as shown in Figure~\ref{fig:study1-time_correctness}.




We found a significant effect of display condition on \textbf{Tree Counting Accuracy} (where lower values are more accurate), per Table~\ref{tab:study1_performance}. \hl{Contrast between Desktop ($M = 6.36$, $SD = 9.25$) and VR ($M = 3.58$, $SD = 4.96$) indicated VR had significantly lower relative error ($t(169) = 2.674$,  \textit{estimate} $= 2.78$, $p = \textbf{0.008}$).}





\subsubsection{Cloud Observation Time}
\begin{figure}
    \centering
    \includegraphics[width=1\linewidth]{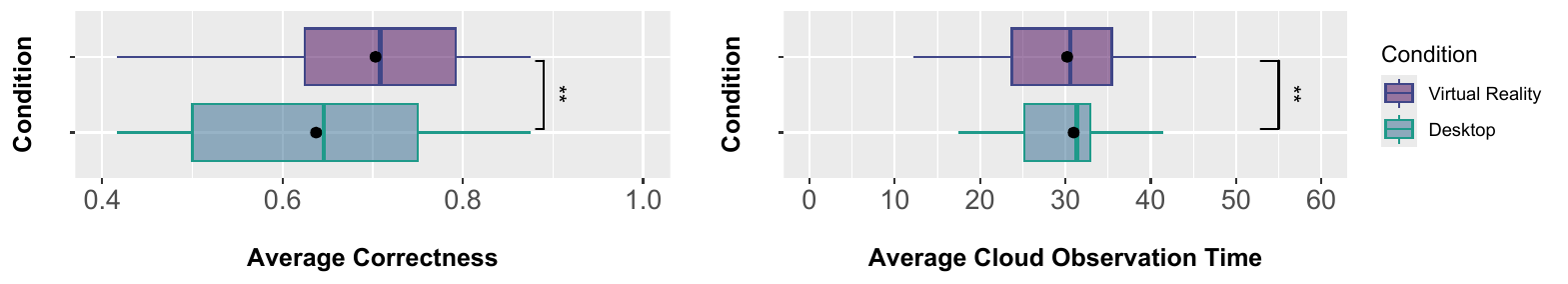}
    \caption{Correctness and cloud observation times. \sigleveltwo}
    \label{fig:study1-time_correctness}
\end{figure}
\begin{figure}
    \centering
    \includegraphics[width=01\linewidth]{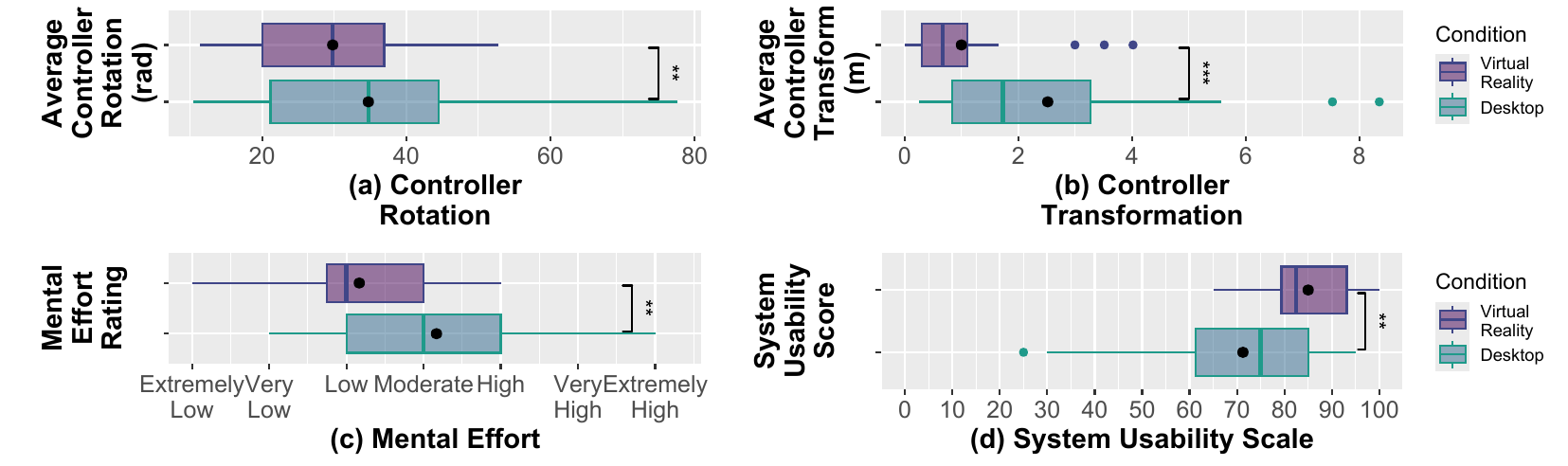}
    \caption{Movement statistics and feedback measures. \sigleveltwo}
    \label{fig:study1-feedback-movement}
    \vspace{-10pt}
\end{figure}
We found a significant effect of display condition on \textbf{Cloud Observation Time}, per Table~\ref{tab:study1_performance}. 
Contrast between Desktop ($M = 37.1$, $SD = 44.5$) and VR ($M = 32.1$, $SD = 30.9$) indicated that Desktop was significantly slower ($t(1139) = 2.582$, \textit{estimate} $= 4.97$, $p = $\textbf{.010}), see Figure~\ref{fig:study1-time_correctness}.

\subsubsection{Frame-of-Reference Transformation}

We found significant effects of display condition on cumulative \textbf{Frame-of-Reference Translation and Rotation}, per Table~\ref{tab:study1_performance} and illustrated in Figure~\ref{fig:study1-feedback-movement}a-b. Contrast between Desktop ($M = 2.25m$, $SD = 3.32$) and VR ($M = 0.99m$, $SD = 1.02$) indicated significantly higher use of translation controls on the Desktop ($t(1139) = 7.351$, $\text{estimate = 0.283}$, $p = \text{\textbf{$<$.001}}$).  Contrast between Desktop ($M = 0.964$, $SD = 0.860$) and VR ($M = 0.4116$, $SD = 0.597$) indicated significantly higher use of rotation controls on the Desktop ($t(1139) = 2.649$, \textit{estimate} $= 4.96$, $p =$ \text{\textbf{0.008}}).

\subsubsection{Participant Feedback}

We found a significant difference in \textbf{Mental Effort} between conditions, as in \appendixtag\ Table~4 
and Figure~\ref{fig:study1-feedback-movement}c. 
Desktop ($M = 4.32$, $SD = 1.22$) needed more effort than VR ($M = 3.29$, $SD = 1.24$). 

Analysis found a similar difference in \textbf{System Usability} scores between conditions, shown in \appendixtag\ Table~4 
and illustrated in Figure~\ref{fig:study1-feedback-movement}d. VR ($M = 83.6$, $SD = 11.6$) scored higher then than desktop ($M = 71.8$, $SD = 17.8$).
The VR and Desktop displays achieved `Excellent' and `Good' usability levels respectively.

Analysis on \textbf{Preference} showed participants preferred VR over desktop displays; 23/24 preferred VR for enjoyment, for viewing point clouds, and expressed an overall preference for VR ($\chi^2 (1,24) = 20.167$, $p =\textbf{$<$.001}$ in each instance); and 20/24 preferred VR for perceived accuracy ($\chi^2 (1,24) = 10.667$, $p = \textbf{0.001}$).



\newcolumntype{M}[1]{>{\raggedright\arraybackslash}m{#1}}

\begin{table}
\centering

\caption{Summary of deviance tables for user performance variables. Significant values highlighted in bold.}
\label{tab:study1_performance}
\small
\begin{tabular}{{M{2.5cm}M{2.2cm}rrr}}
  \hline
 & & $x^2$ & $df$ & $\text{p-value}$ \\ 
  \hline
  \multirow{3}{=}{Cloud Observation Time} & Condition & 6.73 & 1 & \textbf{0.010} \\ 
  & QuestionType & 217.71 & 5 & \textbf{$<$.001} \\ 
  & Condition:QuestionType & 9.61 & 5 & 0.087 \\ 
   \hline
  \multirow{3}{=}{Correctness} & Condition & 6.765 & 1 & \textbf{0.009} \\ 
  & QuestionType & 159.82 & 5 & \textbf{$<$.001} \\ 
  & Condition:QuestionType & 2.792 & 5 & 0.732 \\ 
   \hline
  \multirow{1}{=}{Tree Counting Accuracy} & Condition & \hl{7.192} & 1 & \textbf{0.007} \\
   \hline
  \multirow{3}{=}{Controller Rotation} & Condition & 7.09 & 1 & \textbf{0.008}\\
  & QuestionType & 169.73 & 5 & \textbf{$<$.001} \\ 
  & Condition:QuestionType & 3.60 & 5 & 0.609 \\ 
   \hline
  \multirow{3}{=}{Controller Transformation} & Condition & 54.57 & 1 &\textbf{$<$.001}\\
  & QuestionType & 122.92 & 5 & \textbf{$<$.001} \\ 
  & Condition:QuestionType & 8.00 & 5 & 0.156 \\ 
   \hline
\end{tabular}
\vspace{-10pt}
\end{table}

\subsection{Discussion of Study 1 Results}

VR was found to outperform the Desktop in all three performance metrics, supporting hypothesis \textbf{H1}.
Participants were more likely to determine the correct answer to tasks, perform faster, and achieve better accuracy in the \textit{counting} task.
The depth perception afforded to VR from its stereoscopic view and structure-from-motion outperforms the manual manipulation required by Desktop displays.

Participants greatly preferred the VR display over the Desktop display, supporting hypothesis \textbf{H2}.
VR was the majority choice for every feedback preference category, in addition to showing higher system usability and lower mental effort.
This showed that VR was simply easier and more enjoyable to use for the assigned tasks.

Despite finding that VR improved overall performance, we found no interaction effect of VR for any of the particular question types, contradicting hypothesis \textbf{H3}.
We had expected that question types like split identification would be more strongly effected by being displayed in 3D.
The advantages gained by use of VR is generally consistent for all task types, leading us to the conclusion that it is the best choice for dense forest point cloud analysis.
Ultimately, VR appears to be faster, more accurate, and more user friendly for assessing forest point cloud data than the equivalent Desktop display within the parameters of our assessment. 

The formative study had some limitations. Our participant gender ratio was not balanced. 
Using a controller does not leverage the full utility of VR.
One participant commented that ``it would feel more natural in VR to have hand interactions''.
Matching the resolution of the headset precluded us from testing higher resolution desktop displays.


\section{Focus+Context Techniques for Point Cloud Forest Assessment}
\label{techniques}

To address the challenges of dense point cloud analysis in VR, we designed a set of Focus+Context (F+C) visualisation techniques to support forestry assessment tasks. 
These techniques are informed by \hl{observations and comments from the formative user study} and forestry domain expert feedback.
In this section, we identify the requirements for a functional F+C technique, provide an overview of our visualisation design, and describe our implementation.

\subsection{Requirements}

We conducted a series of interviews with forestry domain experts to capture appropriate requirements. 
This involved informal discussions with four inventory and resource management specialists to identify shortfalls in 2D point cloud visualisation software, and inform the needs for the visualisations.
The discussions helped outline that the techniques need to work with large datasets, on the \textbf{scale} (\textbf{R1}) of compartmental LiDAR scans.
The visualisations need to be of sufficient \textbf{quality} (\textbf{R2}) that it is possible to perform tree feature identification. 
The techniques should be of sufficient \textbf{performance} (\textbf{R3}) that they do not cause discomfort to users.
Additional requirements were determined from feedback from the formative study (Section \ref{study1}).
Users need be able to \textbf{annotate} (\textbf{R4}) the data, visually marking trees of interest.
The technique should provide an indication or \textbf{trace} (\textbf{R5}) of which trees have already been assessed. 
To alleviate obfuscation, users should be able to quickly and responsively \textbf{control} (\textbf{R6}) how much of the data they can see. 
The dataset should have transformation tools, enabling users to change the scale and rotation of the visualisation \textbf{frame-of-reference} (\textbf{R7}). 

\subsection{Visualisation Design}

We use a Focus+Context approach for our visualisation techniques.
Understanding spatial position within a dataset is important for forestry analysis, as the condition of nearby trees is as important as the tree being assessed. 
F+C is a good fit for such analysis, as it allows users to view a scan in high detail, while retaining geospatial context.
The techniques work on the basis of a forest point cloud dataset being represented in a frame-of-reference in two ways:
An abstracted \textit{context} version of the dataset is displayed; the full resolution dataset is aligned to the context, but only displayed within a defined \textit{focus}. 
In this section we describe the components of a F+C technique, then describe our specific solutions in Section \ref{techniquesdesc}.

\subsubsection{Frame-of-Reference}

The Frame-of-Reference (FoR) defines the position and projection of data in the virtual environment, comprising both the context and focus. The user can manipulate the visualisation FoR independent of the user's position, 
adjusting the scale and rotation (\textbf{R7}) using the controller joystick.

\subsubsection{Context}
\label{context}

Our context is modelled as low-resolution fitted cylinders that represent estimated stem points at fixed height intervals.
These cylindrical models provide a simplified dataset that shows the approximate shape, size and location of all the tree stems in the scan without extraneous detail like small branches or leaf matter.
This model (illustrated in Figure {\ref{fig:hero}}b) provides an overview of the full dataset with minimal performance overhead.
The fitted cylinders, though simple, provide feature cues for analysts, letting them see at a glance the approximate structure of any tree.

There are a number of fair use tree/forest LiDAR processing tools available capable of generating such datasets. 
To create our context dataset, we leveraged an AI structural segmentation tool: Forest Structural Complexity Tool (FSCT) by Krisanksi et al.{~\cite{krisanski_forest_2021}}.
\hl{We scanned a forest using an Emesent Hovermap ST} \cite{wilde_emesent_2022} \hl{and processed the scan using FSCT under default parameters, which produced the dataset for the context of our techniques.}
Though the AI models are still improving, the fitted cylinders approach is sufficient to provide geospatial continuity and context to users. 
As these tools develop, there is utility in F+C techniques for validation.

\subsubsection{Focus}

The techniques we describe are distinguished by the area of the focus and how it is controlled. The focus displays a full resolution point cloud (\textbf{R2}) in an area defined by a \textit{control method}, \textit{point of focus}, \textit{shape}, and \textit{size}. The control method defines the position of the point of focus (\textbf{R6}). We explore two control methods: \textit{egocentric} control, where the focus is controlled relative to the user's point-of-view, and \textit{exocentric} control, where controlling the focus is relative to the FoR. For example, a typical egocentric control may be a raycast, whereas exocentric might be a joystick control.

Any shape that can be described in a shader could reasonably be used for the focus.
Users can dynamically adjust the width of the focus area within defined limits that ensure the number of rendered points never exceeds the performance threshold (\textbf{R3}).
This allows users to interact with large point clouds (\textbf{R1}) without harming performance in VR.
The system tracks which trees have been displayed in the focus and applies a subtle colour tint to their context representations, allowing users to see at a glance which sections of the data they have not assessed yet (\textbf{R5}).
\hl{Tree annotation (\textbf{R4}), specifically the selection of individual trees, is supported by applying a colour multiplier to trees as they are selected.}

\subsubsection{Techniques}
\label{techniquesdesc}
We initially experimented with a `flashlight' technique, in which a cylindrical focus extended from the users controller, reminiscent of using a real flashlight to illuminate the data.
This suited fine analysis, but made it challenging to assess tree stems, as users needed to pan up and down repeatedly to ascertain the full structure.
To alleviate this, we projected the focus from the forest floor to the top of the canopy, ensuring the full height of the tree is always visible. 
Through further experimentation we found that a cylindrical focus was ideal for stem assessment, as it fit the shape of the trees, and did not cause obfuscation when assessing trees away from the user.

\begin{table}[]
\centering
\begin{tabular}{cc}
\includegraphics[width=1\linewidth]{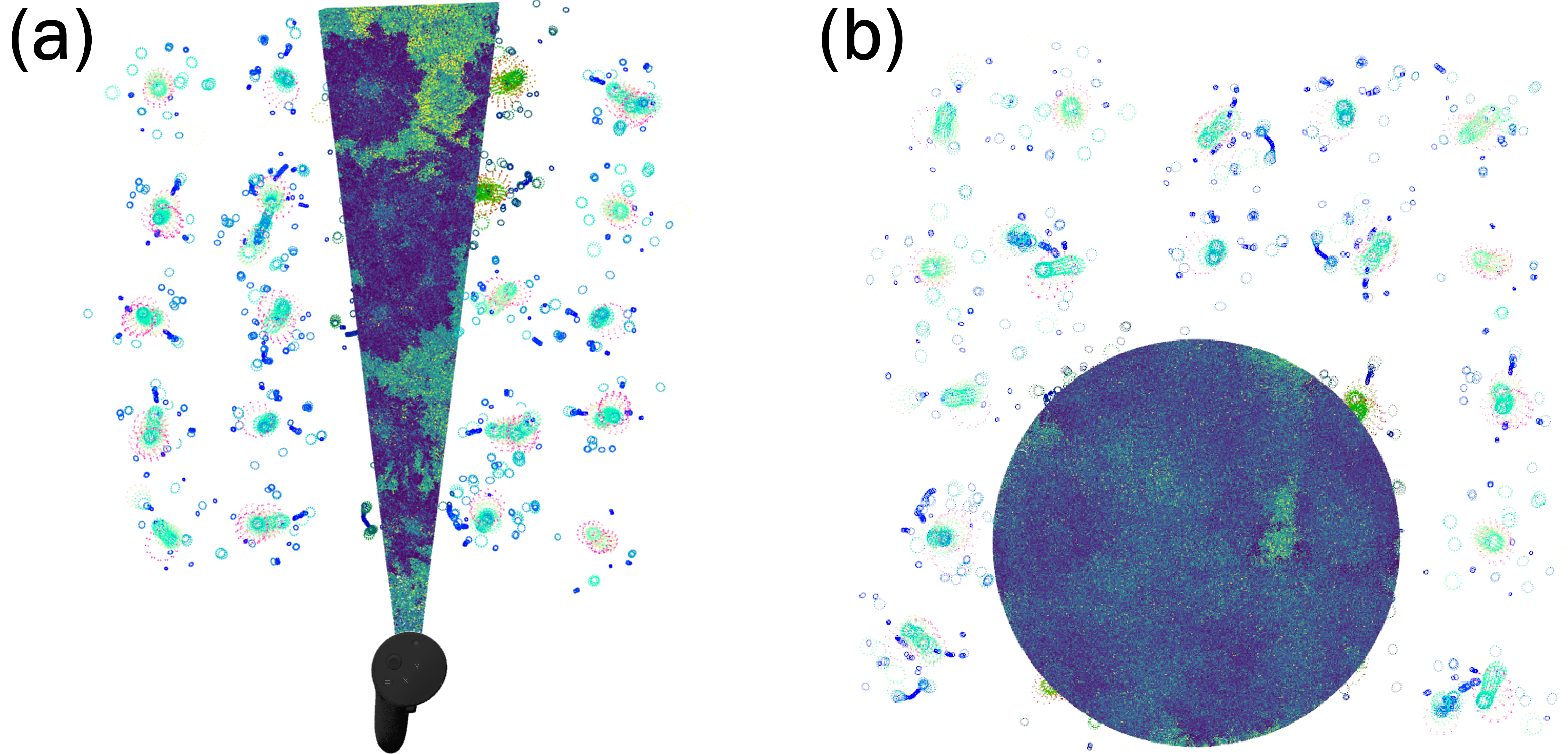}
\end{tabular}
\captionof{figure}{Top view for \hl{(a)} Wedgelight and \hl{(b)} Lantern techniques.}
\label{tab:technique_persepctives}
\vspace{-15pt}
\end{table}

The \textbf{\userfocustag} (egocentric) technique utilises a wedge shaped focus projected from the user's controller and constrained to a two-dimensional orientation parallel to the forest floor.
This results in `wedge' focus that can be pointed wherever the user wants to inspect in detail, shown in Figures \ref{fig:hero}d and \ref{tab:technique_persepctives}a.
Users can control the aperture and focal length of the focus using the controller joystick.

The \textbf{\remotefocustag} (exocentric) technique uses a cylindrical focus at a fixed position, see Figures \ref{fig:hero}c and \ref{tab:technique_persepctives}b. 
The controller joystick is used to move the focus relative to the FoR, and users can independently control its radius with the controller buttons.
    
The \textbf{\rayfocustag} (egocentric) technique uses the same focus shape \hl{and visualisation} as the \remotefocustag\ \hl{but the focus follows the intersection of an infinite raycast from the user's controller to a plane on the forest floor.}
Users point their controller (casting a ray) to aim the focus, \hl{or press a button to pause it.}


Each focus was encapsulated by a transparent mesh as a visual cue to scaling.
This mesh briefly appears visible when the user adjusts scale, then gradually fades back to transparent, as a tangible way to illustrate the focus area to the user.
Points are rendered using an accessible colour scheme \cite{garnier_span_2024}. 
Scale is discrete for the Lantern techniques, and continuous for the \userfocustag. 
Quick-reset buttons were implemented to reset the position of the \remotefocustag, and the scale and rotation of the frame-of-reference.
These features were informed by observations and feedback from Study 1.

\subsection{Implementation}
\label{data-collection-section}

Our techniques are implemented in Unity game engine 2021.3.16f1.
The techniques utilise forestry point clouds in a .las \cite{library_of_congress_laser_2020} file containing a high resolution point cloud of the forest plot.
For the purposes of our study, a LiDAR scan of a production \textit{Pinus radiata} \footnote{Data is publicly available from \url{https://osf.io/hqc9u}.} forest in the Green Triangle region of South Australia was captured
the data from which is shown in Figure \ref{fig:hero}a.




The context data is processed using FSCT as described in Section \ref{context}. Once the focus and context point clouds are imported into Unity, Point Cloud Tools \cite{mika_unitycoderunitypointcloudviewer_2022} is utilised to convert each cloud into a mesh.
A custom material shader is then used to selectively filter the mesh.
The shader takes as input the transformation matrix of the current focus object.
It then calculates if each vertex within the tree mesh is inside or outside of the focus area, storing this result in the vertex's UV coordinates system.
The fragment shader then assesses the values stored in the UVs and discards fragments of the context mesh or focus mesh accordingly.
This results in the full point cloud being drawn within the focus area, and the \contextdatasetdesc{} data being drawn outside of it, as illustrated Figure \ref{fig:hero}c.
A different version of this material exists for each of the selection states of the cloud, allowing subsets (in our case, individual trees) to be drawn in different colours reflecting their state.


\section{Study 2: VR F+C Techniques}

Following our formative study (Section \ref{study1}) and design of our F+C techniques (Section \ref{techniques}), we conducted a summative user study to examine the designed techniques.
The study design was approved by the University Human Research Ethics Committee. 

\subsubsection{Task}

The study task was designed as an analogue for a real world forestry process.
Participants were tasked with assessing forest grids in the virtual environment and marking the trees that displayed the `double leader' characteristic (trees in which the stem splits into two trunks near the base) by touching them with the controller and pressing the trigger, which would \hl{mark} that tree.
Participants were asked to distinguish between single instances of these trees (individuals) and adjacent pairs of these trees (clusters) by \hl{marking} the latter cases an additional time, and asked to identify as many of these individuals and clusters as possible within a sixty second time limit.
This task is a simplified version of the tree selection process described in the introduction \hl{(henceforth \textit{tree marking} for clarity)}.



\subsubsection{Study 2 Dataset Generation}

A forest generator tool was created to synthesise appropriate and realistic datasets for a controlled repeated study of this nature.
This generator used real-world LIDAR scans of 14 single trees: 10 single leaders and 4 double leaders.
The data also included 2 false negative and 2 false positive versions (where focus did not match context) to be representative of a real dataset.
The generator used spatial Poisson disk sampling~\cite{cook_stochastic_1986} to distribute double leader trees around a grid of ordinary trees, resulting in  $1\times10^{228}$ 
possible permutations for each task.
The generator also utilised random scale, rotation and position jitter to create the illusion of a real forest.

\subsubsection{Control}

We designed a Control condition to emulate the constraints of using no F+C technique. 
In the control, the user views either the full context data or the full focus data at any one time, toggled with a button press.
In order to function at an acceptable performance level, the control condition was forced to utilise a lower resolution source point cloud (averaging 50k vertices per tree instead of 500k). 
This concession was deemed necessary as using the full resolution cloud was impossible and independently segmenting the cloud both was insufficient.
Rendering the entire grid at full resolution caused the average framerate to drop below acceptable levels. 
From pilot benchmarking, we found the number of points that could be displayed on our hardware without a drop below 70~fps was 15M, or a maximum of 33 displayable trees.
To mitigate this, the large grid condition was fixed to $15\times15$ trees across all conditions, and the source point cloud was subsampled with a minimum distance of \SI{0.06}{\metre} between points for the control.
This ensured performance never fell below 70~fps.
As an alternative for the control method, we explored but eventually rejected subdividing the plot into sections and loading them individually. Loading and rendering the full-resolution sections would give the participants less than seven seconds per block of 33 trees for our intended task, and would make the assessment of adjacent trees between blocks impossible. 



\subsubsection{Independent Variables}

Independent variables were counterbalanced to avoid effects of fatigue or learning. 
The two independent variables for the study are:

\textbf{Technique:} \userfocustag, \remotefocustag, \rayfocustag, Control. 

\textbf{Grid Size:} Alternate forest grid sizes comprised of rows of trees. 


\subsubsection{Dependent Variables}

The following are the eight independent variables for the study:

\textbf{Tree Identification and Error:} The number of correctly identified double leader trees, with false positives counted as errors.

\textbf{Cluster Identification:} The number of double leaders correctly identified as being individuals or clusters.

\textbf{User Movement:} Cumulative user movement metrics for the user controllers (distance, rotation), headset (distance, rotation), eye movement (rotation), each recorded at intervals of 0.1s.

\textbf{Focus Movement and Scaling:} Cumulative movement metrics for the focus (distance, scale), recorded at intervals of 0.1s.

\textbf{Frame-of-reference Transformation:} Translation and scale metrics of the visualisation frame recorded at intervals of 0.1s.

\textbf{Effort:} Participant rated levels of physical and mental effort after each condition on respective nine point Likert scales.

\textbf{System Usability:} Participants SUS \cite{brooke_sus_1995} questionnaire ratings.

\textbf{Preference:} Participant technique preference for a variety of criteria: accuracy, assessment, enjoyment, and overall favorite.

\subsubsection{Hypothesis}

The hypotheses for the study are as follows:

\textbf{H1} The \textit{\userfocustag, \remotefocustag} and \textit{\rayfocustag} techniques will improve user performance compared to the \textit{Control}.

\textbf{H2} Grid size will effect technique performance.

\textbf{H3} Users will prefer a F+C technique for tree \hl{marking}.

\textbf{H4} A ranking of effectiveness will emerge for the techniques.

\subsubsection{Apparatus}

The study was conducted standing in an unobstructed 
\qtyproduct{3 x 3}{\metre\squared} area to give users freedom of movement. The VR headset and tethered laptop setup described in Section \ref{s1-apparatus} were used for this study.

\subsubsection{Procedure}

The study introduction involved \hl{a pre-study questionnaire and explanation of procedure.}
Participants were acclimatised to the VR headset through fit adjustment and eye tracking setup, then by induction to the study environment.
\hl{ For each condition, they were taught to use the controls, then assessed.}
Training was conducted on a forest grid comprised of $5\times5$ trees with example double leaders, repeatable until the participant was comfortable with the controls. 
The assessment consisted of five large grid tasks ($15\times15$ trees), and five small grid tasks ($10\times10$ trees) for each technique, in random order, with an option to rest between each task and a post condition survey.
After all four conditions were completed, the participant filled out a final post-study questionnaire. This resulted in 2 (forest grid size) $\times$ 5 (repetitions) $\times$ 4 (techniques).

\subsubsection{Participants}

A total 24 participants were recruited from university staff and students (16 male, 8 female) aged from 20--45 ($M=28$, $SD=6.1$). Participants were required to have normal or corrected to normal vision.
Seventeen participants identified as computer scientists, the remainder were divided across psychology, architecture, aviation, education, environmental science (n = 1 each), or unspecified (n = 2). 
\hl{Participants received a \$20AUD ($\approx$\$13USD) gift card.}

\subsection{Study 2 Results}

As in the first study, linear mixed effects
models were employed in R, with fixed effects of interaction technique and grid size, to measures collected across each condition.
Participant variation was accounted for as a random effect. 
The statistical significance of the effects was assessed using Type II Wald Chi-square tests, with a predetermined alpha level of 0.05.
Where grid size was found to be a sole significant factor it was disregarded, as tree and cluster identification percentages were treated as the fraction of the whole, and the different grid sizes had different whole totals. 
The Kruskal-Wallis test was utilised to analyse participant feedback metrics.
Analysis of technique preference was conducted using Chi-square goodness-of-fit tests.
In our analysis, we found no interaction effect between grid size and any other factor on any of our measures.

\newcolumntype{M}[1]{>{\raggedright\arraybackslash}m{#1}}
\begin{table}
\centering
\caption{Summary of deviance tables for user performance variables. Significant values highlighted in bold.}
\label{tab:study2_performance_control}
\small
\begin{tabular}{lrrr}
  \hline
  Response & \( x^2 \) & df & p-value \\ 
  \hline
  Identified Trees & 481.273 & 3 & \textbf{$<$.001} \\ 
  Identified Ratio & 530.518 & 3 & \textbf{$<$.001} \\
  Correctly Marked Trees & 333.116 & 3 & \textbf{$<$.001} \\
  Incorrectly Tagged Trees & 306.973 & 3 & \textbf{$<$.001} \\
  Headset Transformation & 107.249 & 3 & \textbf{$<$.001} \\
  Headset Rotation & 227.026 & 3 & \textbf{$<$.001} \\
  Eye Rotation & 222.168 & 3 & \textbf{$<$.001} \\
  Focus Transformation & 471.692 & 2 & \textbf{$<$.001} \\
  Focus Scale & 558.958 & 2 & \textbf{$<$.001} \\
  Context Rotation & 97.333 & 3 & \textbf{$<$.001} \\
  Context Scale & 74.032 & 3 & \textbf{$<$.001} \\
  \hline
\end{tabular}
\vspace{-5pt}
\end{table}

\subsubsection{Tree Identification and Error}

\begin{figure}
    \centering
    \includegraphics[width=1\linewidth]{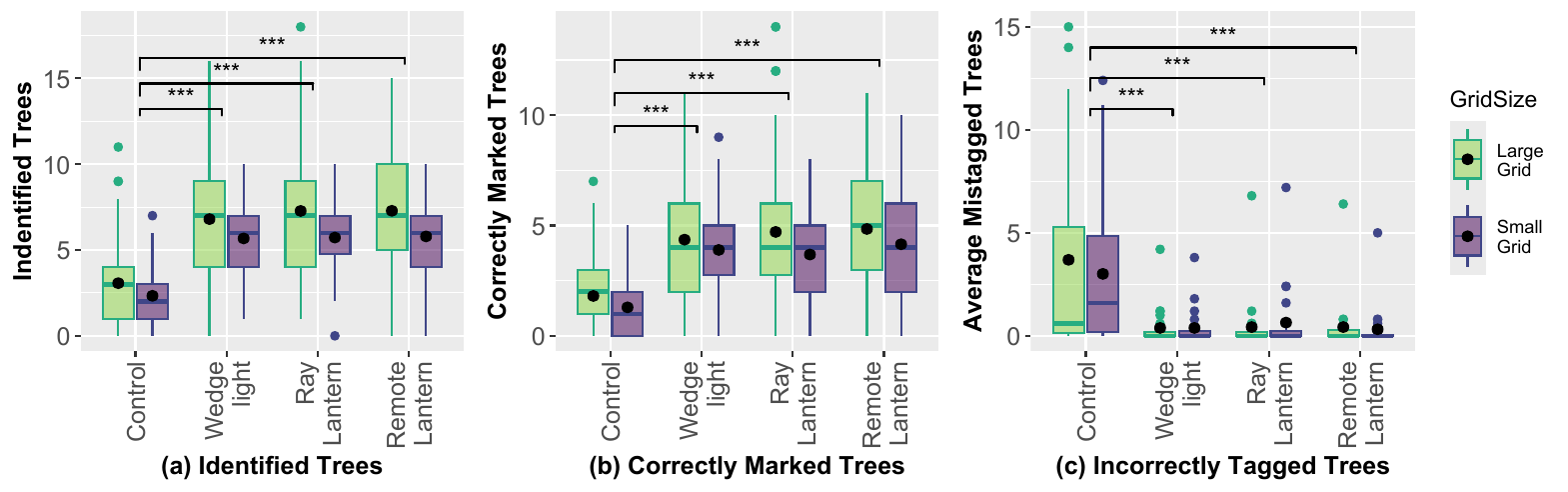}
    \caption{Participant performance metrics. \siglevelthree}
    \label{fig:study2-performance-plot}
    \vspace{-10pt}
\end{figure}

We found a significant effect of interaction technique on \textbf{Identified Tree Number}, per Table~\ref{tab:study2_performance_control}.
Contrasts show the Control ($M = 2.70, SD = 1.97$) to be significantly less effective than \userfocustag\ ($M = 6.24, SD = 2.95$), \rayfocustag\ ($M = 6.5, SD = 2.90$), and \remotefocustag\ ($M = 6.54, SD = 2.85$) for identifying double leader trees as illustrated in Figure~\ref{fig:study2-performance-plot} and \appendixtag\ Table~10. 


Interaction technique had a significant effect on \textbf{Error} per Table~\ref{tab:study2_performance_control} and Figure~\ref{fig:study2-performance-plot}. 
Contrasts between techniques show the Control condition ($M = 3.35, SD = 4.79$) had significantly higher rate of incorrect tree tagging than \userfocustag\ ($M = 0.383$, $SD = 1.10$), \rayfocustag\ ($M = 0.533$, $SD = 1.65$), and \remotefocustag\ ($M = 0.371$, $SD = 1.29$), evidenced in \appendixtag\ Table~10.

Interaction technique had a significant impact on \textbf{Cluster Identification} (Table~\ref{tab:study2_performance_control}). 
Contrasts show the Control ($M = 1.55$, $SD = 1.45$) to be significantly less effective for correctly identifying tree clusters than \userfocustag\ ($M = 4.12$, $SD = 2.37$), \rayfocustag\ ($M = 4.20$, $SD = 2.53$), and \remotefocustag\ ($M = 4.50$, $SD = 2.49$), evidenced in \appendixtag\ Table~10 
and shown in Figure~\ref{fig:study2-performance-plot}. 

\subsubsection{Participant Movement}

\begin{figure}
    \centering
    \includegraphics[width=1\linewidth]{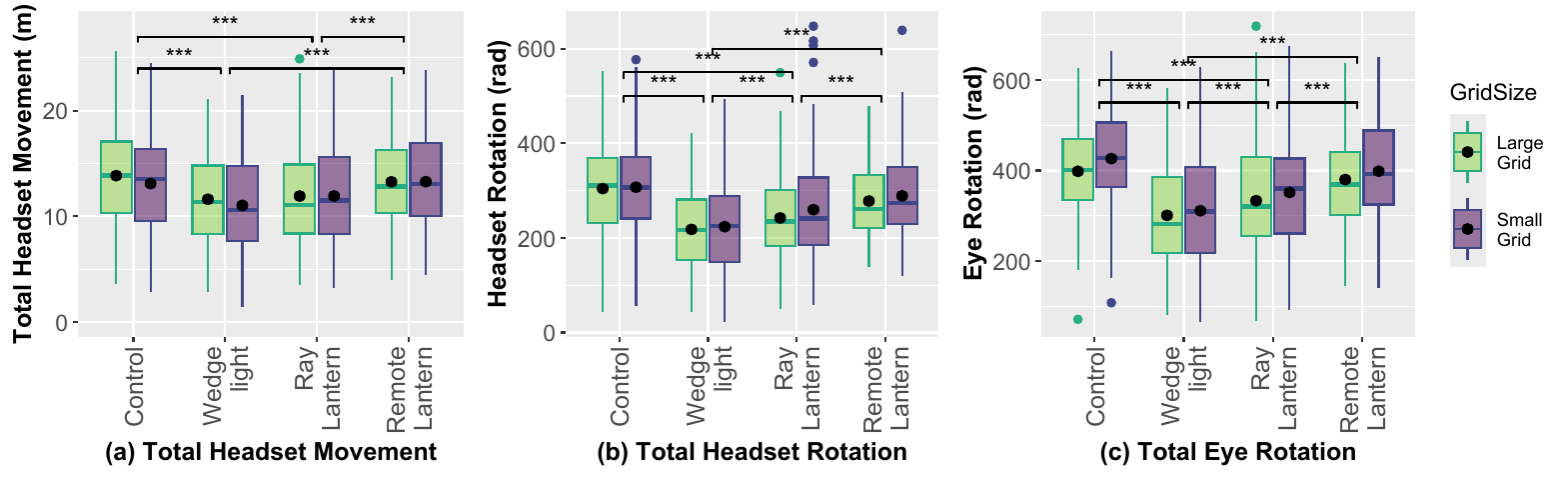}
    \caption{Participant movement metrics. \siglevelthree}
    \label{fig:study2-participant-movement}
    \vspace{-10pt}
\end{figure}


Headset movement, headset rotation, and eye rotation results are shown in Table~\ref{tab:study2_performance_control} and illustrated in Figure \ref{fig:study2-participant-movement}.
We saw a significant effect of interaction technique on cumulative \textbf{Headset Movement}. 
Post-hoc analysis (\appendixtag\ Table~10
) showed the exocentric Control ($M = 13.5m$, $SD = 4.71$) and \remotefocustag\ ($M = 13.3m$, $SD = 4.44$) had significantly higher movement totals than the egocentric \userfocustag\ ($M = 11.3m$, $SD = 4.72$) and \rayfocustag\ ($M = 11.9m$, $SD = 4.57$).

We observed a significant effect of interaction technique on cumulative \textbf{Headset Rotation}. 
Contrasts showed significant differences between each pair of techniques, with the exception of the exocentric techniques Control ($M = 305.7 rad$, $SD = 93.1$) and \remotefocustag\ ($M = 283.4 rad$, $SD = 86.3$). Both of these techniques exhibited significantly more headset rotation than \rayfocustag\ ($M = 250.9 rad$, $SD = 103.0$), and \userfocustag\ ($M = 220.8 rad$, $SD = 91.9$) techniques as evidenced in \appendixtag\ Table~10.

We found a significant effect of interaction technique on cumulative \textbf{Eye Rotation}.
Control condition ($M = 412.2 rad$, $SD = 104.3$) had the the highest eye rotation, followed by \remotefocustag\ ($M = 389.0 rad$, $SD = 107.0$), \rayfocustag\ ($M = 342.3 rad$, $SD = 124.4$), and \userfocustag\ ($M = 306.0 rad$, $SD = 115.0$).
Subsequent analysis revealed that this differentiation was significant between each technique, as evidenced in \appendixtag\ Table~10.

\subsubsection{Interaction Technique Usage}


\begin{figure}
    \centering
    \includegraphics[width=1\linewidth]{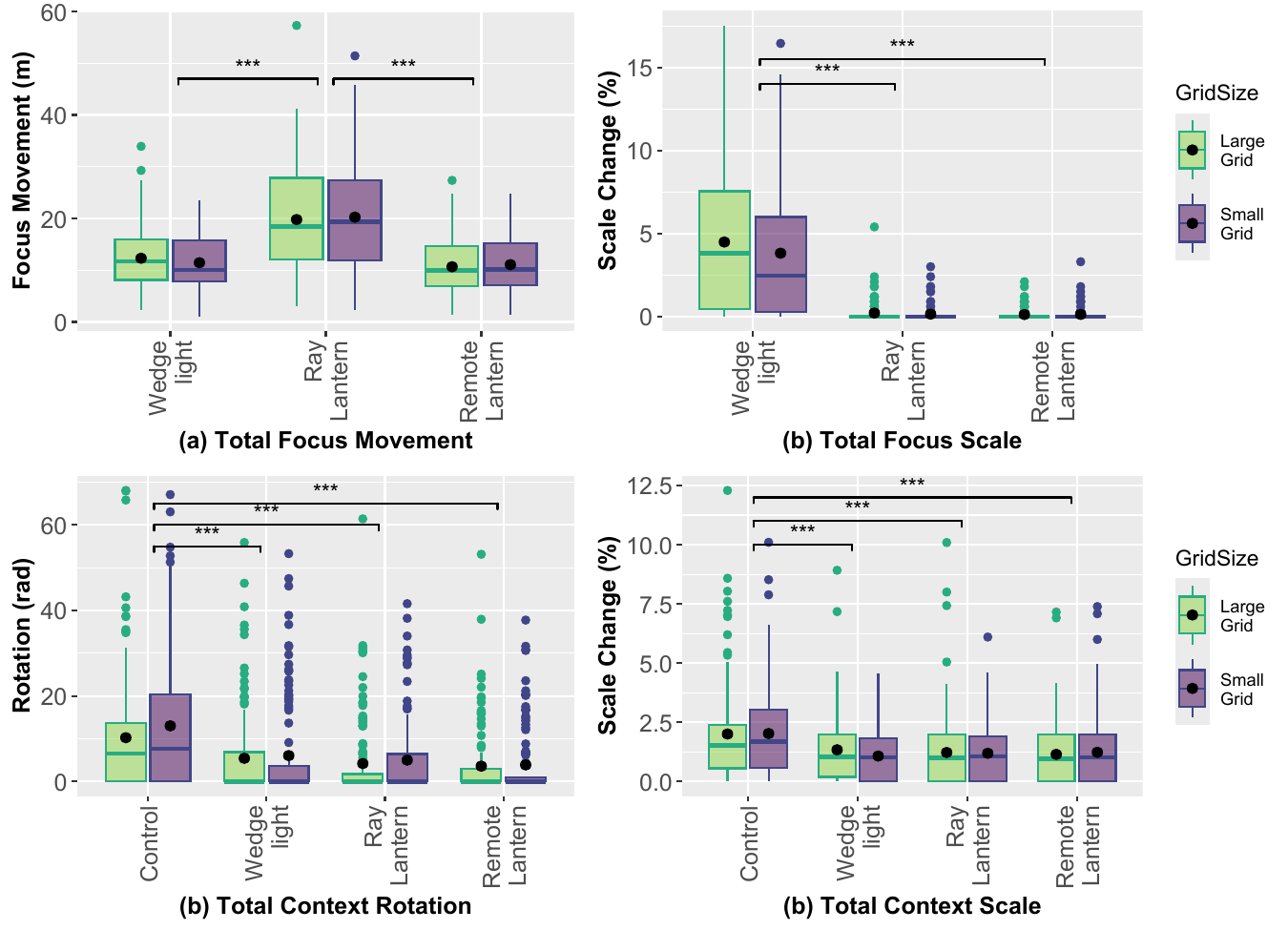}
    \caption{Interaction technique metrics. \siglevelthree}
    \label{fig:study2-interaction-movment}
    \vspace{-10pt}
\end{figure}


We found a significant effect of technique on \textbf{Focus Movement}, per Table~\ref{tab:study2_performance_control} and Figure~\ref{fig:study2-interaction-movment}. 
Contrasts between the techniques in Appendix 1 Table~11 
show \userfocustag\ ($M = 11.9 m$, $SD = 5.61$) and \remotefocustag\ ($M = 10.9 m$, $SD = 5.84$) required significantly less focus movement than \rayfocustag\ ($M = 20.0$, $SD = 9.53$). 

Interaction technique had an additional significant impact on \textbf{Focus Scale}, per Table~\ref{tab:study2_performance_control} and Figure~\ref{fig:study2-interaction-movment}. 
\userfocustag\ ($M = 4.150$, $SD = 4.10$) had a significantly higher level of amount of scale change than either \rayfocustag\ ($M = 0.177$, $SD = 0.57$) or \remotefocustag\, as evidenced in  Appendix 1 Table~11. 

We found significant impact of interaction technique on \textbf{Frame-of-Reference Rotation and Scale}, per Table~\ref{tab:study2_performance_control} and illustrated in Figure~\ref{fig:study2-interaction-movment}. 
Control condition ($M = 11.6 rad$, $SD = 15.2$) had significantly higher FoR rotation than \userfocustag\ ($M = 5.75 rad$, $SD = 11.4$), \rayfocustag\ ($M = 4.61 rad$, $SD = 9.44$), and \remotefocustag\ ($M = 3.74 rad$, $SD = 8.09$), as exhibited in contrasts in Appendix 1 Table 10. 
Similarly, FoR Scale in Control ($M = 2.02$, $SD = 2.03$) was higher than \userfocustag\ ($M = 1.21, SD = 1.23$), \rayfocustag\ ($M = 1.21$, $SD = 1.38$), and \remotefocustag\ ($M = 1.19$, $SD = 1.37$), as evidenced in the contrasts in Appendix 1 Table~10.

\begin{figure}
    \centering
    \includegraphics[width=1\linewidth]{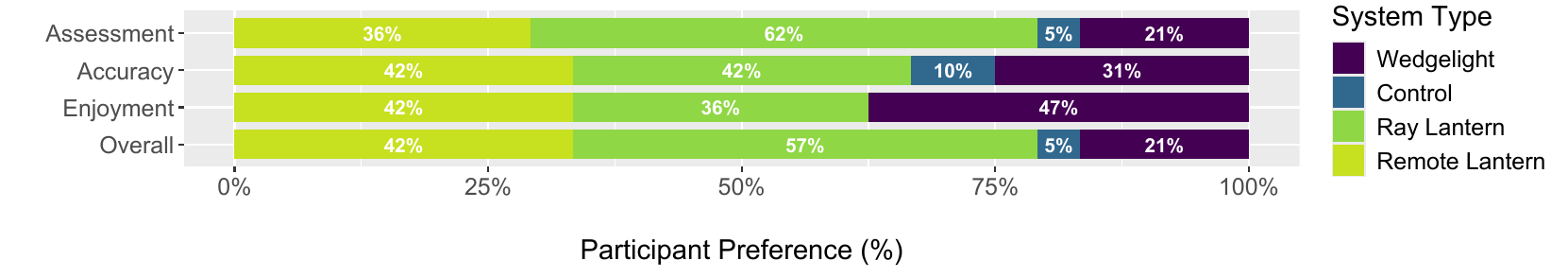}
    \caption{Distribution of participant preferences.}
    \label{fig:participant-preference}
\end{figure}



\subsection{Participant Feedback} 

\begin{figure}
    \centering
    \includegraphics[width=1\linewidth]{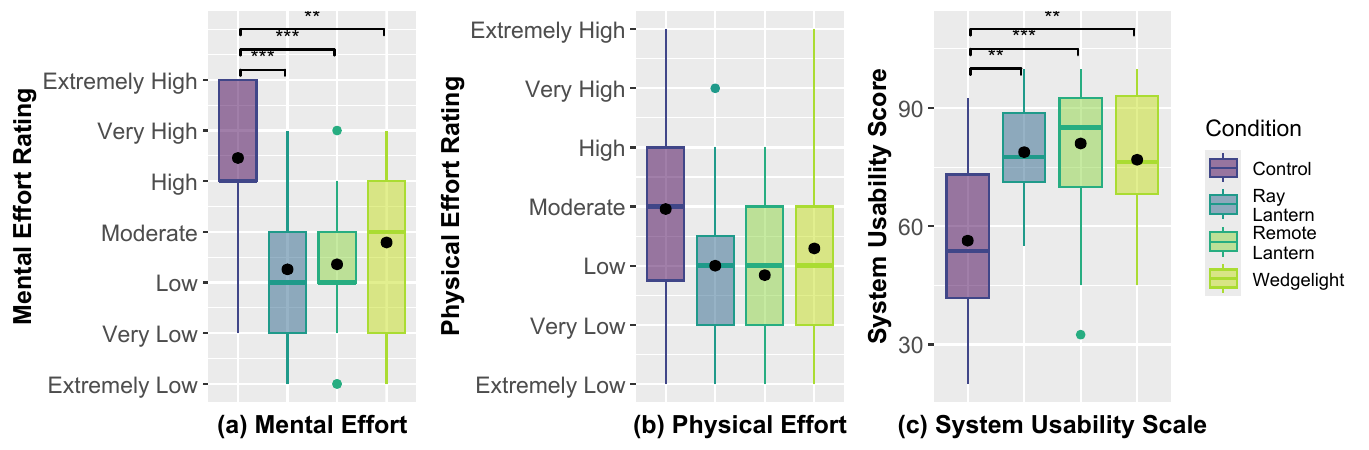}
    \caption{Feedback measures. \sigleveltwo \siglevelthree}
    \label{fig:study2-participant-feedback}
    \vspace{-15pt}
\end{figure}

Interaction technique had no effect on \textbf{Physical Effort} ($\chi^2 (3) = 5.887$, $p = 0.117$), evidenced in Figure~\ref{fig:study2-participant-feedback} and \appendixtag\ Table~7.
Analysis on \textbf{Mental Effort} found a statistically significant difference between the techniques ($\chi^2 (3) = 27.286$, $p =\textbf{$<$.001}$), as illustrated in Figure~\ref{fig:study2-participant-feedback}, and \appendixtag\ Table~7. 
Control ($M = 5.46$, $SD = 1.32$) required higher mental effort than \userfocustag\ ($M = 3.79$, $SD = 1.61$), \rayfocustag\ ($M = 3.26$, $SD = 1.39$) and \remotefocustag\ ($M = 3.36$, $SD = 1.22$), suggesting the techniques demand less mental effort. 

We found a statistically significant difference between the \textbf{System Usability} scores for the different interaction techniques ($\chi^2 (3) = 19.342$, $p =\textbf{$<$.001}$), shown in Figure~\ref{fig:study2-participant-feedback} and \appendixtag\ Table~7. 
\userfocustag\ ($M = 76.9$, $SD = 17.9$), \rayfocustag\ ($M = 78.8$, $SD = 14.8$) and \remotefocustag\ ($M = 81.0$, $SD = 17.5$) achieved levels of `Good', `Good' and `Excellent' respectively.
Each technique was scored higher overall than the Control ($M = 56.4, SD = 21.4$), which was placed in the `Poor' category. 

\textbf{Technique Preference} yielded statistically significant results for all categories, illustrated in Figure~\ref{fig:participant-preference} and detailed in \appendixtag\ Tables~8 and 9. 
Post hoc comparisons using Bonferroni adjusted alpha levels 
indicated that for overall preference, Control was selected substantially less than \rayfocustag\ ($p$ = \textbf{0.004}) and \remotefocustag\ ($p$ = \textbf{$<$0.001}).
For assessment preference, \rayfocustag\ was preferred over the control ($p$ = \textbf{$<$0.001}).
For enjoyment, \remotefocustag\ and \userfocustag\ were preferred over the control ($p$ = \textbf{0.004}, $p$ = \textbf{0.002} respectively).
For accuracy, \remotefocustag\ and \rayfocustag\ were preferred over the Control ($p$ = \textbf{0.004} in both cases).

\subsection{Discussion of Study 2 Results}

The F+C techniques were found to outperform the \controltag\ in every performance metric, supporting hypothesis \textbf{H1}.
The techniques granted users significantly higher rates of successful tree identification, tree \hl{marking}, and overall accuracy. 
The most significant impact of the techniques was the lower rate of errors, and hence increase in overall accuracy rate. 
This suggests the techniques are advantageous for preventing false positives in tree identification.

We found no evidence that the size of grid used for the task effected the performance of any F+C technique, contradicting hypothesis \textbf{H2}.
As in real operations, the total area of the forest is less important for assessment than the proportion of trees being thinned.

Users overwhelmingly preferred the F+C techniques for tree \hl{marking} over the \controltag\, supporting hypothesis \textbf{H3}.
This is unsurprising, as reported mental effort required was significantly higher for the \controltag, and it was the only condition for which system usability score fell below the `good' range.
The cylindrical focus shape for \rayfocustag\ and \remotefocustag\ made them well suited to the study task.
This is likely why they were ranked first and second in overall preference.
The \userfocustag\ technique was less suited to the task, but was rated highest for enjoyment.
We expect it would excel at finer analysis of tree structure. 
We did not find clear rankings between the F+C techniques (failing to support \textbf{H4}). Our findings suggest that F+C in general is a good method for presenting complex point cloud data.

Participant comments provided insight to how the techniques might be improved, such as the option to freely change the shape of the focus (e.g., square, circle, column), more options for highlighting trees when illuminated (e.g., set a time threshold for the colour change), and expand on the controls, specifically: raising and lowering of the context, individually rotating each tree around its axis, and dynamically scale up trees in focus.

The techniques are transferable to other fields, though alternate methods of creating context are needed for other domains.
They could be used in construction geometry quality inspection, as viewing the source data circumvents the minimum fidelity problem~{\cite{wang_applications_2019}}.
They could also be used to validate outputs of data abstraction processes, like those applied in archaeological artifact analysis~{\cite{poux_3d_2017}} and agricultural structure detection~{\cite{comba_unsupervised_2018}}.

Our study was not without limitations.
The primary constraint was the need to use lower resolution data for the \controltag\ condition.
When the performance of VR is on par with the fidelity of remote sensing, the techniques should be compared directly.
The gender ratio of participants was not evenly balanced.
To improve relevance to real world forest management, it should be expanded to other task analogues, such as tree species and age classes.


\section{Conclusions}

Our formative user study showed users were faster and more accurate using VR than desktop for forestry analytics tasks.
The depth perception and freedom of movement offered by VR displays make them ideal for structural analysis of dense point clouds, even requiring less controller movement than on the desktop.
VR has inherent advantages for forestry data analytics.

Subsequently we described a series of F+C techniques, leveraging these advantages for dense forest point cloud assessment.
Our follow up user study showed these techniques all performed equally better than the control, leading us to conclude that F+C techniques are a natural fit for this branch of forestry analytics.
We believe they are viable for other fields with similar spatial data, and we have shared our study data, point clouds and scripts for future research.

The cylindrical focus methodology described for Lantern techniques best suits comparative analysis of multiple trees, whereas the directed focus for the Wedgelight technique would likely be more useful for more detailed structural analytics.
This implementation is not the only way to design F+C techniques for these tasks; however, future research should explore other methods for generating context data.
The ability to switch between multiple diverse synthetic datasets would be an interesting way to explore the interaction between AI data analysis and human understanding.
Virtualised tree analysis in the future would benefit from implementation of F+C tools for point cloud analysis.

\acknowledgments{%
	\hl{This work was supported in part by a Australian Government Research Training Program Scholarship, OneFortyOne Plantations Pty Ltd, Forestry Australia, and the Gottstein Trust.}%
}

\bibliographystyle{abbrv-doi}

\bibliography{references_fixed}

\clearpage

\end{document}